# Bose-Einstein condensation in a gas of the Fibonacci oscillators

## Abdullah Algin[*]

Department of Physics, Eskisehir Osmangazi University, Meselik, 26480-Eskisehir, Turkey

**Abstract**

We consider a system of the two-parameter deformed boson oscillators whose spectrum is given by a generalized Fibonacci sequence. In order to obtain the role of the deformation parameters $(q_1, q_2)$ on the thermostatistics of the system, we calculate several thermostatistical functions in the thermodynamical limit and investigate the low-temperature behavior of the system. In this framework, we show that the thermostatistics of the $(q_1, q_2)$-bosons can be studied by the formalism of Fibonacci calculus which generalizes the recently proposed formalism of $q$-calculus. We also discuss the conditions under which the Bose-Einstein condensation would occur in the present two-parameter generalized boson gas. However, the ordinary boson gas results can be obtained by applying the limit $q_1 = q_2 = 1$.

---

[*] E-mail address: aalgin@ogu.edu.tr

## 1. Introduction

The role of the quantum harmonic oscillator in physics, especially its deformation with some parameter, is crucial. It is often used in the investiga tion of a variety of physical systems from quantum optics [1] to angular momentum realizations. In the past decade, deformed algebras have most popularly found applications in field and string theories [2]. A realization of the quantum algebra $su_q(2)$ using a $q$-analogue of the usual bosonic harmonic oscillator and the Jordan -Schwinger mapping has been developed by Biedenharn [3] and Macfarlane [4]. In particular, the intimate relation between $q$-oscillators and quantum groups [5 -7] has been extensively investigated [8 -10].

On the other hand, another direction of applications has focused on the field of statistical mechanics. There are two distinct methods in the literature for studying the generalized statistical mechanics. The first method i s to use one or two -parameter deformed bosonic and fermionic quantum algebras. The second method is the formalism of Tsallis nonextensive statistical mechanics. In this context, possible connections between quantum groups and Tsallis nonextensive statistic al mechanics have been investigated [11-14].

In the framework of $q$-bosons and similar operators the so called quons [15], considerable efforts have been made for obtaining possible violation of the Pauli exclusion principle [16] and also possible relation to anyonic statistics [17,18]. However, it was shown in [19-25] that the high- and low-temperature behaviors of the quantum group symmetric bosonic and fermionic oscillator models depend radically on the real deformation parameters. Although other two -parameter realizations have been studied in the literature [26-28], a complete formalism for the generalization of the conventional physical quantities of the bosons and fermions coming out by deformations is still under investigation.

The aim of this paper is two-fold: First, we wish to study the low -temperature thermostatistical properties in a gas of the two -parameter deformed bosonic oscillators whose spectrum is given by a generalized Fibonacci sequence. Second, we want to show that the thermostatistics of these $(q_1, q_2)$-bosons can be studied by the formalism of Fibonacci calculus which generalizes the earlier one -parameter deformed formalism called $q$-calculus [29-35]. In this sense, we continue the works of Lavagno and Narayana Swamy [31,32], and consider a two-parameter generalization of bosonic system, which is called the Fibonacci oscillators. By means of the formalism of Fibonacci calculus, we particularly study the thermostatistics of a gas of the commuting Fibonacci oscillators. We obtain the low-temperature thermostatistical properties of such two-parameter boson model through a two-parameter generalized statistical distribution function of the system. We also discuss the effect of the real independent deformation parameters $q_1$ and $q_2$ on the conditions under which the Bose-Einstein condensation would occur.

Beside the present application of the Fibonacci oscillators to the statistical mechanics, now we would like to present the main reasons why one may consider two distinct deformation parameters $(q_1, q_2)$ in other physical applications. The reasons can be separated into two groups as follows:

The first one is related to quantum algebraical properties: (i) The Fibonacci oscillators offer a unification of quantum oscillators related to quantum groups [36,37]. (ii) The covariant Fibonacci oscillator algebra is the most general quantum group invariant bosonic oscillator algebra. The invariance quantum group of this oscillator is the $SU_r(d)$ with $r = q_1/q_2$. In this sense, if the quantum group symmetry is preserved, then the number of deformation parameters in $d$ dimensions should be just two [38]. This is an important property in order to have a connection with quantum group formalism. (iii) Although there are some studies showing a close connection between relativity, $q$-oscillators and difference operators [39,40], the multidimensional Fibonacci oscillator corresponding to the two-parameter basic number definition can be interpreted as a relativistic oscillator corresponding to the bound state of two bosonic particles with masses $m_1$ and $m_2$ [36,37]. Therefore, the additional parameter $q_2$ has a physical significance so that it can be related to the mass of the second bosonic particle in a two-particle relativistic quantum harmonic oscillator bound state [36]. (iv) In our recent study [25], we showed that the covariant Fibonacci oscillator gas model with $SU_{q_1/q_2}(2)$-symmetry exhibits the Bose-Einstein condensation in a specific interval of the deformation parameters $q_1$ and $q_2$. But this model has a different Hamiltonian from the present commuting Fibonacci oscillator gas, which will be defined in section 3. Strictly speaking, the $SU_{q_1/q_2}(2)$-covariant boson model Hamiltonian proposed in [21,25] was constructed a new and different representation for the covariant Fibonacci oscillator operators. The two-parameter $SU_{q_1/q_2}(2)$-covariant boson gas model exhibits an interesting behaviour for high temperatures [21] such that its willingness of fermion-like behaviour increases too much for a wide range of the deformation parameters $q_1$ and $q_2$. Furthermore, the fermionic version of the present Fibonacci oscillators introduced in [41] has revealed some notable properties for high temperatures [23]. Such a two-parameter fermion model in two spatial dimensions exhibits remarkably anionic type of behaviour at some critical values of the model deformation parameters. However, it is impossible to obtain a similar behavior neither in the one-parameter deformed $SU_q(2)$ fermion gas model [19] nor in the undeformed fermion gas. From such interesting results, we show that the $(q_1, q_2)$-deformed Fibonacci oscillator algebra is a new candidate to study systems with fractional statistics. (vi) Some integrable models and solvable lattice models have benefited from the properties of two parameter quantum algebras. For instance, it was shown in [42] that the quantum algebraic structure of the spin-1/2 XXZ chain with twisted periodic boundary conditions is a two-parameter deformed algebra $SU_{q,t}(2)$.

On the other hand, the second group of reasons comes originally from the phenomenological studies [43,44]: (i) The quantum algebra with two deformation parameters may have more flexibility when dealing with application to the concrete physical models [45]. Although any quantum algebra with one or more deformation parameters may be mapped onto the standard single-parameter case [46,47], the physical results obtained from a $(p,q)$-deformed oscillator system are not the same. In particular, it is recently argued that the $(p,q)$-deformed phonon systems may be useful and effective in order to deal with anharmonicity or/and interactions of phonons in the realistic condensed matter [27,28]. (ii) The recent results arising from the application of the $qp$-rotator model with $U_{qp}(u_2)$-symmetry to the description of rotational bands of various atomic nuclei show also the importance for introducing the second deformation parameter. In this sense, the results obtained from the $qp$-rotator model are better than the ones derived from the $q$-rotator model and the $\kappa$-Poincare model [48,49]. In such an application, the two deformation parameters of the $qp$-rotator model had an interpretation as inertial parameters that describe the softness of the deformed and superdeformed nuclei. (iii) Recently, quantum algebra with one deformation parameter has been used in phenomenological description of particle properties [50,51]. In particular, $q$-bosons have been used to describe unusual behaviour of the intercept (or the strength) $\lambda$ of the two-particle correlation function. This technique gives a direct correspondence between the deformation parameter $q$ and the intercept $\lambda$. For the higher order correlations, the theories of two deformation parameters become important. It is argued in [51] that the two-parameter deformed bosonic oscillator is a candidate to describe an explicit form of the intercepts $\lambda^{(n)}$ of $n$-particle correlation functions with $n \geq 3$ of identical pions or kaons. However, there is a remarkable difference between the covariant Fibonacci oscillators and the so called $q$-bosons. Strictly speaking, the $q$-bosons [3,4,52] do not have a covariance under the action of the quantum group $SU_q(d)$. (iv) Recently, the one-parameter deformed fermion algebra with $SU_q(2)$-symmetry has been used to discuss some hadronic properties such as the dynamical mass generated for the quarks and the pure nuclear pairing force version of the BCS many-body formalism [53,54]. Also, it has been used to understand higher order effects in the many-body interactions in nuclei [55,56]. In this framework, possible applications of a fermionic version [41] of the present two-parameter Fibonacci oscillator algebra could provide new results to understand interactions between fermions and bosons. Such studies would provide us to understand the physical meaning of the deformation parameters more precisely.

Thus, the present two-parameter bosonic Fibonacci oscillator model and its fermionic version in [41] may be used to approximate physical systems similar to the ones mentioned above. Also, all above considerations give the reasons to consider two distinct deformation parameters in physical appli cations, and therefore, show the importance of two-parameter deformed oscillator algebra studies.

The paper is organized as follows : In section 2, we review the general properties of the bosonic Fibonacci oscillator algebra and specialize for the commuting Fibonacci oscillator system. In section 3, we investigate the low-temperature thermostatistical properties of the commuting Fibonacci oscillator gas in the thermodynamical limit . The distribution function and other thermostatistical functions of the syste m are derived in terms of two deformation parameters via the grand partition function of the system. Particular emphasis is given to a discussion of the Bose-Einstein condensation phenomenon for the commuting Fibonacci oscillator gas. In the last section, we give our conclusions.

## 2. The Fibonacci oscillator algebra

The $d$-dimensional usual boson oscillators satisfy the following commutation relations:

$$\left[b_i, b_j^*\right] = \delta_{ij}, \qquad i, j = 1, 2, ...., d,$$

$$\left[b_i, b_j\right] = \left[b_i^*, b_j^*\right] = 0, \tag{1}$$

$$b_i^* b_i = N_i, \qquad b_i b_i^* = N_i + 1,$$

where $b_i$ and $b_i^*$ are the bosonic annihilation and creation operators, respectively, and $N_i$ is the boson number operator. These oscillators are invariant under SU($d$) transformations. The most general deformation of the above al gebra [36,37] will be of the form

$$a^* a = [N], \qquad aa^* = [N+1], \tag{2}$$

where $[n]$ is a generalized number. Imposing the cond itions

$$[0] = 0, \qquad a|0\rangle = 0, \tag{3}$$

the eigenvalues of $[N] = a^* a$ are given by the basic integers $[0], [1], [2],...$ In terms of the above representation, the commutation relations of the earliest one-dimensional $q$-oscillator [57] is given by

$$aa^* - q^2 a^* a = 1, \qquad 0 < q < 1,$$

$$[a, N] = a, \qquad [a^*, N] = -a^*. \tag{4}$$

The $q$-commutation relation in Eq. (4) can be rewritten as

$$[N+1] = q^2 [N] + 1, \tag{5}$$

where the number operator spectrum of $[N]$ was defined by the relation

$$[n] = \frac{1 - q^{2n}}{1 - q^2}, \tag{6}$$

which yields the Jackson basic integers [29].

On the other hand, the another $q$-oscillator algebra associated with the symmetrized basic integer

$$[n] = \frac{q^{2n} - q^{-2n}}{q^2 - q^{-2}}, \tag{7}$$

obeys the commutation relation [3,4]

$$aa^* - q^2 a^* a = q^{-2N}. \tag{8}$$

This commutation relation can be alternatively rewritten as [3 6]

$$[N+2] = (q^2 + q^{-2})[N+1] - [N], \tag{9}$$

together with the initial conditions $[0] = 0, [1] = 1$. We also note that the inhomogeneous difference equation given in Eq. (5) can be expressed as a second order hom ogeneous difference equation

$$[N+2] = (q^2 + 1)[N+1] - q^2 [N]. \tag{10}$$

Therefore, the difference equations corresponding to the two types of $q$-deformations considered above are special cases of the general linear second order homogeneous difference equation [3 6]

$$[N+2] = \alpha [N+1] + \beta [N], \tag{11}$$

which upon choice of initial conditions used in Eq. (9) generates the generalized Fibonacci basic integer

$$[n] = \frac{q_2^{2n} - q_1^{2n}}{q_2^2 - q_1^2} , \tag{12}$$

with $\alpha = q_1^2 + q_2^2$, $\beta = -q_1^2 q_2^2$. Here $q_1$ and $q_2$ are real independent deformation parameters. We note that the basic number definitions in Eqs. (6) and (7) are special cases of Eq. (12). The Fibonacci oscillator having the spectrum given by Eq. (1 2) can also be expressed as the operator algebra

$$aa^* - q_1^2 a^* a = q_2^{2N} ,$$

$$aa^* - q_2^2 a^* a = q_1^{2N} . \tag{13}$$

The generalization of this system into $d$-dimensional case can be done by considering $d$-commuting copies of the creation and annihilation operators of the one - dimensional two-parameter deformed oscillator algebra defined by Eq. (13) [3 6,38]. Thus, the $d$-dimensional commuting Fibonacci oscillator algebra can be written in terms of the following relations [36]:

$$[a_i, a_j] = 0 , \qquad i, j = 1,2,.....,d,$$

$$[a_i, a_j^*] = 0 , \qquad i \neq j ,$$

$$a_i a_i^* - q_1^2 a_i^* a_i = q_2^{2N_i} ,$$

$$a_i a_i^* - q_2^2 a_i^* a_i = q_1^{2N_i} , \tag{14}$$

$$a_i^* a_i = [N_i], \qquad a_i a_i^* = [N_i + 1],$$

where the spectrum of the deformed boson number operators $[N_i]$ is defined by the generalized Fibonacci basic integers in Eq. (12). We also note that the Fibonacci oscillator algebra has symmetry under the interchange of the deformation parameters $q_1$ and $q_2$.

Another way of getting algebra for the $d$-dimensional Fibonacci oscillator is to construct the covariant Fibonacci oscillators which have the same bosonic degeneracy as the usual boson oscillators in Eq. (1). In this way, the quantum group covaria nt two-parameter deformed bosonic oscillator algebra can be obtained [3 6] as follows:

$$c_i c_k = q_1 q_2^{-1} c_k c_i, \qquad i < k,$$

$$c_i c_k^* = q_1 q_2 c_k^* c_i, \qquad i \neq k,$$

$$c_1 c_1^* - q_1^2 c_1^* c_1 = q_2^{2N},$$

$$c_k c_k^* - q_1^2 c_k^* c_k = c_{k-1} c_{k-1}^* - q_2^2 c_{k-1}^* c_{k-1}, \quad k = 2,......d,$$

$$q_1^{2N} = c_d c_d^* - q_2^2 c_d^* c_d,$$

(15)

where the total deformed boson number operator for this system is

$$c_1^* c_1 + c_2^* c_2 + ........ + c_d^* c_d = \left[ N_1 + ........ + N_d \right] = \left[ N \right],$$

whose spectrum is given by the generalized Fibonacci basic integers $\left[ n \right]$ in Eq. (12). Such an algebra was recently used to investigate the thermodynamical properties of a gas of the quantum group covariant oscillators in the high and low -temperature limits [21,24,25].

Thus, the Fibonacci oscillator offers a unification of oscillators related to quantum groups [26,36,37]. There are two types of the multidimensional Fibonacci oscillators: commuting and covariant. The two types are related by a transformation and the diagonal commutation relations for both types of oscillators are the same. The commuting Fibonacci oscillators are relevant for the construction of coherent states and for constructing unitary quantum Lie algebras, whereas the covariant Fibonacci oscillators are needed for quantum group covariance and a bilinear Hamiltonian with a degenerate spectrum [3 6]. For both types of Fibonacci oscillators, the usual boson oscillator algebra in Eq. (1) can be recovered in the limit $q_1 = q_2 = 1$.

Now we would like to give the basic properties of the Fibonacci calculus through the concept of coherent states [3 6]. The commuting Fibonacci oscillators in

Eqs. (14) possess coherent states which are eigenvectors of the annihilation operators $a_i$ as

$$a_i |z_1, z_2, ....., z_d\rangle = z_i |z_1, z_2, ....., z_d\rangle,$$
(16)

$$|z_1, z_2, ....., z_d\rangle = \sum_{n_i \geq 0} \frac{z_1^{n_1} .....z_d^{n_d}}{[n_1]! .....[n_d]!} (a_1^*)^{n_1} .....(a_d^*)^{n_d} |0,...,0\rangle,$$

where we have defined $(a_1^*)^{n_1} .....(a_d^*)^{n_d} |0,...,0\rangle$ as $|n_1, n_2, ..., n_d\rangle$ and $[n_i]$ is given by Eq. (12). Using the coherent states, one can show that to every vector $|f\rangle$ of the Hilbert space there corresponds an analytic function given by

$$\langle \bar{z}_1, \bar{z}_2, ....., \bar{z}_d |f\rangle = f(z_1, z_2, ....., z_d) = \sum_{n_i \geq 0} f_{n_1, ..., n_d} z_1^{n_1} .....z_d^{n_d} .$$
(17)

The Fibonacci difference operator $\partial_i$, and multiplication by $z_i$ are the counterparts of the $a_i$, $a_i^*$ operators, respectively, and are defined by

$$\langle \bar{z}_1, \bar{z}_2, ....., \bar{z}_d |a_j|f\rangle = \frac{f(z_1, ..., q_1^2 z_j, ..., z_d) - f(z_1, ..., q_2^2 z_j, ..., z_d)}{(q_1^2 - q_2^2) z_j} = \partial_j f(z_1, ..., z_d),$$
(18)

$$\langle \bar{z}_1, \bar{z}_2, ....., \bar{z}_d |a_j^*|f\rangle = z_j f(z_1, ..., z_d).$$
(19)

For the sake of simplicity, we will consider the one-dimensional Fibonacci oscillator, and hence we have the following important relations [3 6]:

$$\langle \bar{z} |a|f\rangle = \partial^{(q_1, q_2)} f(z) = \frac{f(q_1^2 z) - f(q_2^2 z)}{(q_1^2 - q_2^2) z},$$
(20)

$$\langle \bar{z} |a^*|f\rangle = z f(z).$$
(21)

From Eqs. (18)-(21), the transformation from Fock observables to the configuration space can be formally represented as

$$a^* \to z, \qquad a \to \partial^{(q_1, q_2)},$$
(22)

where $\partial^{(q_1, q_2)}$ is the Fibonacci difference operator defined in Eq. (20). The Fibonacci difference operator generalizes its earlier versions connected with the $q$-basic number definitions given in Eqs. (6) and (7) [3,4,29,30,38, 57]. The Jackson derivatives (JD) related with the $q$-basic numbers in Eqs. (6) and (7) can be obtained from the Fibonacci difference operator in Eq. (20) by applying the limits $q_2 = 1$ and $q_2 = q_1^{-1}$, respectively. Obviously, the ordinary derivative can be recovered in the limit $q_1 = q_2 = 1$.

From the above discussion, we should remark that the Fibonacci oscillators share most of nice properties of the $q$-oscillators in Eqs. (4) and (8). The only property which does not hold is the Leibniz rule for the exterior derivative on the Fibonacci-Manin plane [36,37]. We will see below that the above mentioned properties of Fibonacci calculus play a central role for studying the thermostatistics of a gas of the commuting Fibonacci oscillators. In particular, several thermostatistical func tions of the system can be obtained by using the Fibonacci difference operator in Eq. (20).

## 3. Low-temperature thermostatistics of the Fibonacci oscillators

In this section we investigate the low-temperature (high-density) behavior of the commuting Fibonacci oscillators defined in Eq. (14). Unlike to the covariant Fibonacci oscillator gas model studied in [21,25], the system containing the commuting Fibonacci oscillators constitutes essentially a "free" $(q_1, q_2)$-defomed bosonic gas system, s ince the Fibonacci oscillators do not interact with each other. The reason behind this consideration is that we do not have both a specific deformed commutation relation between bosonic annihilation (or creation) operators and a quantum group symmetry structure in Eq. (14).

In the grand canonical ensemble, we choose the Hamiltonian of such a free $(q_1, q_2)$-deformed bosonic Fibonacci oscillator gas as

$$\hat{H}_{q_1, q_2} = \sum_i (\varepsilon_i - \mu_{q_1, q_2}) \hat{N}_i, \tag{23}$$

where $\varepsilon_i$ is the kinetic energy of a particle in the state $i$, $\mu_{q_1, q_2}$ is the $(q_1, q_2)$-deformed chemical potential, and $\hat{N}_i$ is the boson number operator relative to $\varepsilon_i$. Similar Hamiltonians were also considered by several other authors [31 -35,58-75]. It should be noted that this Hamiltonian is a two-parameter deformed Hamiltonian and depends implicitly on the deformation parameters $q_1$ and $q_2$, since the number operator is deformed by means of Eq. (14).

Thermal average of observables can be calculated by using the usual prescription of quantum mechanics. Therefore, the thermal average of an operator is written in the standard form

$$< \hat{O} > = \frac{Tr(\hat{O} e^{-\beta \hat{H}_{q_1, q_2}})}{Z}, \qquad (24)$$

where $Z$ is the grand canonical partition function defined as

$$Z = Tr(e^{-\beta \hat{H}_{q_1, q_2}}), \qquad (25)$$

and $\beta = 1/k_B T$, $k_B$ is the Boltzmann constant, $T$ is the temperature of the system. As in the one-parameter deformed boson gas case [31-35,75], we should remark that the structure of the density matrix $\rho = e^{-\beta \hat{H}_{q_1, q_2}}$ and the thermal average are undeformed. Hence, the structure of the partition function is unchanged. This is a nontrivial assumption, since it implicitly gives an unmodified structur e of the Boltzmann-Gibbs entropy,

$$S_{q_1, q_2} = k_B \ln W_{q_1, q_2}, \qquad (26)$$

where $W_{q_1, q_2}$ stands for the number of states of the system corresponding to the set of occupation numbers $\{f_{i, q_1, q_2}\}$. Obviously, the number $W_{q_1, q_2}$ is modified in the present two-parameter case. It should be emphasized that this is a different deformation from the Tsallis nonextensive statistics [11], where the structure of the entropy is deformed via the logarithmic function.

By following the procedure proposed by [64] for a one -parameter deformed noninteracting boson gas, we now derive the $(q_1, q_2)$-deformed Bose-Einstein distribution function. We define the mean occupation number $f_{i, q_1, q_2}$ corresponding to $\hat{N}_i$ by

$$q_1^{2 f_{i, q_1, q_2}} = \frac{1}{Z} Tr(e^{-\beta \hat{H}_{q_1, q_2}} q_1^{2 \hat{N}_i}), \qquad q_2^{2 f_{i, q_1, q_2}} = \frac{1}{Z} Tr(e^{-\beta \hat{H}_{q_1, q_2}} q_2^{2 \hat{N}_i}). \qquad (27)$$

From Eqs. (27) and (14), we derive

$$[f_{i,q_1,q_2}] = \frac{1}{Z} Tr(e^{-\beta \hat{H}_{q_1,q_2}} [\hat{N}_i]),$$ (28)

which can be alternatively written as

$$[f_{i,q_1,q_2}] = \frac{1}{Z} Tr(e^{-\beta \hat{H}_{q_1,q_2}} a_i^* a_i).$$ (29)

From the cyclic property of the trace [59,60] and using the commuting Fibonacci oscillator algebra in Eq. (14), we obtain

$$\frac{[f_{i,q_1,q_2}]}{[f_{i,q_1,q_2} + 1]} = e^{-\beta (\varepsilon_i - \mu_{q_1,q_2})}.$$ (30)

Using Eqs. (12) and (30), we derive

$$f_{i,q_1,q_2} = \frac{1}{[\ln(q_1^2 / q_2^2)]} \ln \left( \frac{z_{q_1,q_2}^{-1} e^{\beta \varepsilon_i} - q_2^2}{z_{q_1,q_2}^{-1} e^{\beta \varepsilon_i} - q_1^2} \right),$$ (31)

where $z_{q_1,q_2} = e^{\beta \mu_{q_1,q_2}}$ is the $(q_1, q_2)$-deformed fugacity of the system. This equation provides the $(q_1, q_2)$-deformed statistical distribution function of the commuting Fibonacci oscillators. The total number of particles is given by $N = \sum_i f_{i,q_1,q_2}$. The $(q_1, q_2)$-deformed distribution function $f_{i,q_1,q_2}$ possesses the following properties:

(1)    In the limit $q_1 = q_2 = 1$, Eq. (31) reduces to the usual Bose-Einstein distribution.

(2)    The distribution function $f_{i,q_1,q_2}$ should be nonnegative. This results in the following constraints on the $(q_1, q_2)$-deformed fugacity and the chemical potential:

$$z_{q_1,q_2} \leq \begin{cases} q_2^{-4}, & \mu_{q_1,q_2} \leq -4k_B T \ln q_2, & (q_2 > q_1), \\ q_1^{-4}, & \mu_{q_1,q_2} \leq -4k_B T \ln q_1, & (q_2 < q_1). \end{cases}$$ (32)

When we take the limit $q_1 = q_2 = 1$, Eq. (32) reduces to

$$z \equiv z_{1,1} \leq 1 \quad \text{or} \quad \mu \equiv \mu_{1,1} \leq 0,$$ (33)

which are the same constraints in the fugacity and the chemical potential as in the usual boson gas. Also, we note that the $(q_1, q_2)$-deformed fugacity $z_{q_1,q_2}$ is independent of the dimension of the Fibonacci oscillators.

(3)    $f_{i,q_1,q_2}$ possesses the symmetry property such that $f_{i,q_1,q_2} = f_{i,q_2,q_1}$ .

(4)    $f_{i,q_1,q_2}$ has some important limiting cases. In the limit $q_2 = 1$ and $q_1 = q^{1/2}$, one can obtain the one-parameter deformed distribution of $q$-bosons [31,33,76] satisfying the algebra in Eqs. (4)-(6). When we take the limit $q_1 = q^{1/2}$ and $q_2 = q^{-1/2}$, the statistical distribution of $q$-bosons [32,34,35,75] satisfying the another algebra in Eqs. (7)-(9) can also be recovered.

Figures 1 and 2 show the $(q_1, q_2)$-deformed statistical distribution function $f_{i,q_1,q_2}$ for finite temperatures as a function of $\eta = \beta(\varepsilon_i - \mu_{q_1,q_2})$ for several values of the deformation parameters $q_1$ and $q_2$.

From Eqs. (25) and (23), one can deduce the logarithm of the bosonic grand partition function as

$$\ln Z = -\sum_i \ln(1 - z_{q_1,q_2} e^{-\beta\varepsilon_i}). \tag{34}$$

This form is due to the fact that we have chosen the Hamiltonian to be a linear function of the boson number operator but it is not linear in $a_i^* a_i$, which can be deduced from Eq. (14). For this reason, the standard thermodynamic relations in the usual form are ruled out as in the case of the one-parameter deformed boson gas [31-35,75]. For instance, the total number of particles in the commuting Fibonacci oscillator gas model cannot be obtained by using the standard thermodynamical expression such as

$$N \neq z\left(\frac{\partial}{\partial z}\right) \ln Z. \tag{35}$$

Here it is important that the Fibonacci difference operator in Eq. (20) should be used instead of the ordinary thermodynamics derivative with respect to $z$ as follows:

$$\frac{\partial}{\partial z} \to D_z^{(q_1, q_2)},\qquad(36)$$

where $D_z^{(q_1, q_2)}$ may be called as the modified Fibonacci difference operator as

$$D_z^{(q_1, q_2)} = \frac{(q_1^2 - q_2^2)}{\ln(q_1^2 / q_2^2)} \partial_z^{(q_1, q_2)}.\qquad(37)$$

Therefore, the total number of particles in the commuting Fibonacci oscillator gas can be derived from the constraint

$$N = z D_z^{(q_1, q_2)} \ln Z \equiv \sum_i f_{i, q_1, q_2},\qquad(38)$$

where $f_{i, q_1, q_2}$ is expressed by Eq. (31). In order to obtain the low-temperature behavior of the system, we can replace the summations by integrals for a large volume and particle number. However, we note that the $\vec{p} = \vec{0}$ case plays a special role in the ideal Bose gas [77-79]. Since $\ln Z$ diverges in the $\vec{p} = \vec{0}$ term as $z \to 1$, we separately account for the term $\vec{p} = \vec{0}$ as a second term in the following equation of state:

$$\frac{P}{kT} = -\frac{4\pi}{h^3} \int_0^\infty p^2 dp \ln(1 - z_{q_1, q_2} e^{-\beta p^2 / 2m}) - \frac{1}{V} \ln(1 - z_{q_1, q_2}).\qquad(39)$$

Similarly, the particle density for the commuting Fibonacci oscillators is

$$\frac{1}{\upsilon} = \frac{N}{V} = \frac{4\pi}{h^3} \int_0^\infty p^2 dp \frac{1}{\ln(q_1^2 / q_2^2)} \ln\left[ \frac{(z_{q_1, q_2}^{-1} e^{\beta p^2 / 2m} - q_2^2)}{(z_{q_1, q_2}^{-1} e^{\beta p^2 / 2m} - q_1^2)} \right] + \frac{1}{V} \frac{1}{\ln(q_1^2 / q_2^2)} \ln\left[ \frac{(1 - z_{q_1, q_2} q_2^2)}{(1 - z_{q_1, q_2} q_1^2)} \right].\ (40)$$

Eqs. (39) and (40) can be rewritten as

$$\frac{P}{kT} = \frac{1}{\lambda^3} g_{5/2}(q_1, q_2, z_{q_1, q_2}) - \frac{1}{V} \ln(1 - z_{q_1, q_2}),\qquad(41)$$

$$\frac{1}{\upsilon} = \frac{1}{\lambda^3} g_{3/2}(q_1, q_2, z_{q_1, q_2}) + \frac{1}{V} \frac{1}{\ln(q_1^2 / q_2^2)} \ln\left[ \frac{(1 - z_{q_1, q_2} q_2^2)}{(1 - z_{q_1, q_2} q_1^2)} \right],\qquad(42)$$

where $\lambda = \sqrt{2\pi\hbar^2/mkT}$ is the thermal wavelength. The two-parameter generalized Bose-Einstein function $g_n(q_1, q_2, z_{q_1,q_2})$ is defined as follows:

$$g_n(q_1, q_2, z_{q_1,q_2}) = \frac{1}{\Gamma(n)} \int_0^\infty x^{n-1} dx \frac{1}{\ln(q_1^2/q_2^2)} \ln\left[\frac{(z_{q_1,q_2}^{-1} e^x - q_2^2)}{(z_{q_1,q_2}^{-1} e^x - q_1^2)}\right]$$

$$\tag{43}$$

$$= \frac{1}{\ln(q_1^2/q_2^2)}\left(\sum_{l=1}^\infty \frac{(q_1^2 z_{q_1,q_2})^l}{l^{n+1}} - \sum_{l=1}^\infty \frac{(q_2^2 z_{q_1,q_2})^l}{l^{n+1}}\right),$$

where $x^2 = \beta p^2/2m$. This two-parameter generalized function reduces to the standard Bose-Einstein function $g_n(z)$ in the limit $q_1 = q_2 = 1$. Moreover, the limit $q_2 = 1$ and $q_1 = q^{1/2}$ gives the one-parameter deformed $g_n(q, z)$ functions of [31,33]. One can also recover the $q$-deformed $h_n(q, z)$ functions of [32,34,35] in the limit $q_2 = q^{-1/2}$ and $q_1 = q^{1/2}$. However, one should note that the $(q_1, q_2)$-deformed Bose-Einstein functions $g_n(q_1, q_2, z_{q_1,q_2})$ do not satisfy the property $zD_z^{(q_1,q_2)} g_n(q_1, q_2, z_{q_1,q_2}) \neq g_{n-1}(q_1, q_2, z_{q_1,q_2})$ due to the existence of the Fibonacci difference operator $D_z^{(q_1,q_2)}$ in Eq. (43).

In figures 3 and 4, the $(q_1, q_2)$-deformed functions $g_{3/2}(q_1, q_2, z_{q_1,q_2})$ and $g_{5/2}(q_1, q_2, z_{q_1,q_2})$ are shown as a function of $z$ for several values of the deformation parameters $q_1$ and $q_2$ for the cases $(q_1, q_2) \leq 1$ and $(q_1, q_2) \geq 1$, respectively. Here, the upper bound of $z$ is $q_2^{-4}$ for $q_2 > q_1$, and it is $q_1^{-4}$ for $q_2 < q_1$, which were obtained from Eq. (32). When we compare with the $q_1 = q_2 = 1$ case in figures 3 and 4, the values of the $(q_1, q_2)$-deformed Bose-Einstein functions $g_{3/2}(q_1, q_2, z_{q_1,q_2})$ and $g_{5/2}(q_1, q_2, z_{q_1,q_2})$ decrease for $(q_1, q_2) \leq 1$, while they increase for $(q_1, q_2) \geq 1$.

Eqs. (31) or (42) imply that $<n_0>$ is the average occupation number for the zero momentum state

$$<n_0> = \frac{1}{\ln(q_1^2/q_2^2)} \ln\left[\frac{(1-z_{q_1,q_2}q_2^2)}{(1-z_{q_1,q_2}q_1^2)}\right]. \tag{44}$$

This term contributes significantly to Eq. (42) if $<n_0>/V$ is a finite number, i.e., if a finite fraction of the commuting Fibonacci oscillators occupies the single level with $\vec{p} = \vec{0}$. This fact gives rise to the famous phenomena of Bose-Einstein condensation. In this context, one can rewrite Eq. (42) as

$$\frac{\lambda^3 <n_0>}{V} = \frac{\lambda^3}{\upsilon} - g_{3/2}(q_1,q_2,z_{q_1,q_2}), \tag{45}$$

which implies $(<n_0>/V) > 0$ when the critical combination of the temperature and the specific volume occurs such that the $(q_1,q_2)$-deformed fugacity $z_{q_1,q_2}$ will reach its maximum value given in Eq. (32). Therefore, we obtain

$$\frac{\lambda^3}{\upsilon} \geq g_{3/2}(q_1,q_2,z_{q_1,q_2}). \tag{46}$$

This phenomenon is referred to as the Bose-Einstein condensation. The value of the $(q_1,q_2)$-deformed function $g_{3/2}(q_1,q_2,z_{q_1,q_2})$ depends on the deformation parameters $q_1$ and $q_2$. Therefore, these parameters are responsible for the low-temperature behavior of the present commuting Fibonacci oscillator gas model.

The critical temperature $T_c(q_1,q_2)$ for the system can be found from Eq. (46) as

$$T_c(q_1,q_2) = \frac{2\pi\hbar^2/mk}{[\upsilon\, g_{3/2}(q_1,q_2,z_{q_1,q_2})]^{2/3}}. \tag{47}$$

According to figure 3, the critical temperature for the commuting Fibonacci oscillators is much larger than the critical temperature $T_c(1,1)$ for an undeformed boson gas in the special regions of the second deformation parameter $q_2$ close to zero and one. Moreover, one can find a relation between the critical temperature of the commuting Fibonacci oscillator gas and of the undeformed boson gas:

$$\frac{T_c(q_1,q_2)}{T_c(1,1)} = \left(\frac{2.61}{g_{3/2}(q_1,q_2,z_{q_1,q_2})}\right)^{2/3}. \tag{48}$$

In figures 5 and 6, we show the plots of Eq. (48) as a function of the deformation parameters $q_1$ and $q_2$ for the cases $(q_1,q_2) \le 1$ and $(q_1,q_2) \ge 1$, respectively.

The internal energy $U$ of the commuting Fibonacci oscillator gas can be found by extending the procedure proposed in [31-35] to the present two-parameter case. In this calculation, we consider the prescription for the Fibonacci difference operator in Eq. (37) and the ordinary chain rule as follows:

$$U = (-\frac{\partial \ln Z}{\partial \beta}) = \sum_i \frac{\partial y_i}{\partial \beta} D_{y_i}^{(q_1,q_2)} \ln(1 - z_{q_1,q_2} y_i), \tag{49}$$

where $y_i = \exp(-\beta \varepsilon_i)$. This equation leads to

$$U = \sum_i \varepsilon_i f_{i,q_1,q_2}, \tag{50}$$

where $f_{i,q_1,q_2}$ is expressed by Eq. (31). We can also obtain the internal energy as

$$\frac{U}{V} = \frac{3}{2} \frac{kT}{\lambda^3} g_{5/2}(q_1,q_2,z_{q_1,q_2}). \tag{51}$$

With the above results in mind, the specific heat of the commuting Fibonacci oscillator gas can be obtained from the thermodynamic definition $C_V = (\partial U/\partial T)_{V,N}$. For low temperatures, namely in the limit $T < T_c(q_1,q_2)$, the specific heat of our model is

$$\frac{C_V}{Nk} = \frac{15}{4} \frac{\upsilon}{\lambda^3} (z D_z^{(q_1,q_2)} g_{7/2}(q_1,q_2,z_{q_1,q_2})), \tag{52}$$

which can be rewritten in terms of the critical temperature $T_c(q_1,q_2)$ by means of Eq. (47) as

$$\frac{C_V}{Nk} = \frac{15}{4} \frac{(z D_z^{(q_1,q_2)} g_{7/2}(q_1,q_2,z_{q_1,q_2}))}{g_{3/2}(q_1,q_2,z_{q_1,q_2})} \left(\frac{T}{T_c(q_1,q_2)}\right)^{3/2}. \tag{53}$$

On the other hand, the specific heat for the commuting Fibonacci oscillator gas in the limit $T > T_c(q_1,q_2)$ can be approximated as

$$\frac{C_V}{Nk} \approx \frac{3}{2}\frac{(q_1^2 - q_2^2)}{\ln(q_1^2/q_2^2)} + \frac{3}{4.2^{9/2}}\big[2\big](5\big[2\big]+4)\, g_{3/2}(q_1, q_2, z_{q_1, q_2})\left(\frac{T_c(q_1, q_2)}{T}\right)^{3/2}, \qquad (54)$$

where the generalized Fibonacci basic integer [n] is defined in Eq. (12). From Eqs. ( 53) and (54), we deduce the gap in the specific heat in the limit $T = T_c(q_1, q_2)$ as

$$\frac{\Delta C_V}{Nk} \approx \left\{\frac{15}{4}\frac{(z\mathrm{D}_z^{(q_1, q_2)}g_{7/2}(q_1, q_2, z_{q_1, q_2}))}{g_{3/2}(q_1, q_2, z_{q_1, q_2})} - \left[\frac{3}{2}\frac{(q_1^2 - q_2^2)}{\ln(q_1^2/q_2^2)} + \frac{3}{4.2^{9/2}}\big[2\big](5\big[2\big]+4)\, g_{3/2}(q_1, q_2, z_{q_1, q_2})\right]\right\}. (55)$$

In figures 7 and 8, we show the plots of the specific heat $C_V/Nk$ as a function of $T/T_c(q_1, q_2)$ for values of the deformation parameters $q_1$ and $q_2$ for the cases $(q_1, q_2) \le 1$ and $(q_1, q_2) \ge 1$, respectively. In figures 9 and 10, we also show the plots of the gap in specific heat $\Delta C_V/Nk$, using Eq. (55), as a function of the deformation parameters $q_1$ and $q_2$ for the cases $(q_1, q_2) \le 1$ and $(q_1, q_2) \ge 1$, respectively.

Before closing this section, we should emphasize that the results in Eqs. ( 31)-(55) are not only different from the results of the one -parameter boson model studied in [31-35], but also they generalize the results to the case with two deforma tion parameters by means of the Fibonacci oscillators.

By considering the above results, the effect of two deformation parameters on the thermostatistics of the system will be discussed in the next section.

## 4. Discussion and conclusions

In this paper, we studied the behavior of a gas of the commuting Fibonacci oscillators at low temperatures. The Hamiltonian of this system does not show covariance under a quantum group structure. Therefore, the system of the Fibonacci oscillators satisfying algebra in Eq. (14) constitutes essentially an example of non -interacting multi-mode system of the $(q_1, q_2)$ -deformed bosonic particles. The algebra of Fibonacci oscillators has two notable properties: (i) It has a symmetry under the exchange of the defor mation parameters $q_1$ and $q_2$, and (ii) they have values in the interval $0 < (q_1, q_2) < \infty$.

Starting with the commuting Fibonacci oscillators, we also showed that the thermostatistics of the $(q_1, q_2)$ -deformed bosons can be studied by the formalism of Fibonacci calculus, which generalizes the recently proposed formalism of $q$-calculus in [31-35]. The key point to mention is that if we use the modified Fibonacci difference operator in Eq. (37) instead of the ordinary derivatives in the thermodynamical relations, then the standard structure of thermodynamics is not changed even in the present $(q_1, q_2)$ -deformed quantum group approach.

Furthermore, the low-temperature thermostatistical pro perties of the Fibonacci oscillator gas are obtained in the thermodynamical limit. Starting with a $(q_1, q_2)$ -deformed Bose-Einstein distribution function, several thermostatistical functions via the grand partition function of the system are calculated. For instance, the average occupation number, the critical temperature, the internal energy are derived for low temperatures. Subsequently, the specific heat of the system is obtained in the low and high-temperature limits. We then focused on th e effect of the deformation parameters $q_1$ and $q_2$ on these results. We should emphasize that the values of these $(q_1, q_2)$ -deformed functions and thus all other thermostatistical functions change more rapidly to those $q_2$ values that are small deviations of this deformation parameter close to zero and one, respectively.

Moreover, the $(q_1, q_2)$ -deformed Bose-Einstein distribution function in Eq. (31) takes the standard form for an undeformed boson gas in the limit $q_1 = q_2 = 1$. According to figures 1 and 2, the values of the $(q_1, q_2)$ -deformed Bose-Einstein distribution function $f_{i, q_1, q_2}$ for the case $(q_1, q_2) > 1$ increases with the values of the second deformation parameter $q_2$. However, for the case $(q_1, q_2) < 1$, it decreases when the second deformation parameter $q_2$ is decreased.

The specific heat of the commuting Fibona cci oscillator gas shows a discontinuity at the critical temperature as shown in figures 7 and 8. This means that the Bose-Einstein condensation in the commuting Fibonacci oscillator gas is a second -order phase transition. Also, the specific heat of the sy stem has a $\lambda$ -point transition behavior which is not exhibited by the undeformed boson gas. Such a behavior may be physically important in studies on superfluidity. An interesting point is that when the second deformation parameter $q_2$ increases, the discontinuity in the specific heat of the system also increases (figures 7 and 8). However, such a result is in contrast to the behaviour of the $SU_{q_1/q_2}(2)$-covariant boson gas model in [25]. On the other hand, the discontinuity in the specific heat of the system decreases for values of the second deformation parameter $q_2$ close to zero (figure 7). Obviously, in the limit $q_1 = q_2 = 1$, such a discontinuity disappears as in an undeformed boson gas.

Furthermore, the gap in the specific heat of the commuting Fibonacci oscillator gas at the condensation temperature increases with the values of the second deformation parameter $q_2$ (figure 10), and conversely, it decreases for those values of the second deformation parameter $q_2$ close to zero (figure 9). Thus, the system containing commuting Fibonacci oscillators shows the Bose-Einstein condensation for low temperatures in the intervals $0 < (q_1, q_2) < 1$ and $1 < (q_1, q_2) < \infty$. In some sense, the entire behaviour of the system is characterized by the model parameters $q_1$ and $q_2$.

The experimental critical temperature for $^4He$, $T_c \approx 2.18$ K [77], corresponds to the values of deformation parameters $q_1 \approx 1.06$ and $q_2 \approx 1.58$. The same values of deformation parameters are compatible with the gap in the specific heat of a dilute gas of rubidium atoms [80]. The importance of these comparisons lies in the fact that the Fibonacci oscillators algebra can be used to represent some $d$-component interacting systems coupled with the deformation parameters $q_1$ and $q_2$. This consideration was indeed an important notion in our previous works [21,23-25,44], where we used the $SU_{q_1/q_2}(d)$-covariant Fibonacci oscillator algebra defined in Eq. (15). The purpose of these studies was to observe whether an interacting Hamiltonian for a system containing $d$-component bosonic or fermionic particle families can be obtained if one invokes a quantum group symmetry structure to the standard Hamiltonian of the system. However, this does not mean that deformation is equivalent to an interaction. Since quantum deformation may not necessarily be the same as an interaction among the particles of the $d$-component system. The interaction conjectured in the $SU_{q_1/q_2}(d)$- covariant bosonic and fermionic Fibonacci oscillat or gas models comes from the unitary quantum group symmetry of the system s presented by the deformed Hamiltonian s. Furthermore, when we look at the equation of state of these quantum group covariant boson and fermion gas models for the high temperature lim it [21,23,24], all of the virial coefficients except the lowest-order one depend s on the real deformation pa rameters $q_1$ and $q_2$. Therefore, a quantum group symmetric Fibonacci oscillator gas model does incorporate an interaction among its constitu ents. Such an interaction also reflects itself as a deformation in the virial coefficients of the system under consideration. However, we should emphasize that this result not only has a different basis but also contains a different analysis from [76], except that both of the studies pursued the same goal whether there is equivalence between deformation and interaction .

On the other hand, the above comparison may also be physically important, since some recent studies with o ne deformation parameter similarly adduced some values for the deformation parameter [81,82]. Hence, one can also view two parameter deformations as a phenomenological means of introducing extra parameters, " $q_1, q_2$ ", to account for some non-linear properties in the system. Indeed, such an approach with one deformation parameter was considered in [81], where a value of $q$ is found to fit the properties of a real (non -ideal) laser.

As a final remark , it would be interesting to investigate possible implications of the Fibonacci oscillator algebra in the framework of Tsallis nonextensive statistic s. This would provide some new connections between the generalized statistical mechanics and quantum group approach. We hope th at these problems will be addressed in the near future.

**Acknowledgments**

The author wishes to thank the referee for useful suggestions which improved the quality of the paper. This work is supported by the Scientific and Technological Research Council of Turkey (TUBITAK) under the project number 106T593. Also, the author thanks E. Arslan for her help in preparing the three -dimensional plots.

**Figure Captions**

**Figure 1.** The $(q_1, q_2)$-deformed Bose-Einstein distribution $f_{i,q_1,q_2}$ as a function of $\eta = \beta(\varepsilon_i - \mu_{q_1,q_2})$ for various values of the deformation parameters $q_2$ and $q_1 \leq 1$ for finite temperatures.

**Figure 2.** The $(q_1, q_2)$-deformed Bose-Einstein distribution $f_{i,q_1,q_2}$ as a function of $\eta = \beta(\varepsilon_i - \mu_{q_1,q_2})$ for various values of the deformation parameters $q_2$ and $q_1 \geq 1$ for finite temperatures.

**Figure 3.** The $(q_1, q_2)$-deformed Bose-Einstein function $g_{3/2}(z, q_1, q_2)$ as a function of $z$ for the cases $(q_1, q_2) \leq 1$ (above) and $(q_1, q_2) \geq 1$ (below).

**Figure 4.** The $(q_1, q_2)$-deformed Bose-Einstein function $g_{5/2}(z, q_1, q_2)$ as a function of $z$ for the cases $(q_1, q_2) \leq 1$ (above) and $(q_1, q_2) \geq 1$ (below).

**Figure 5.** The ratio $T_c(q_1, q_2)/T_c(1,1)$ of the $(q_1, q_2)$-deformed critical temperature $T_c(q_1, q_2)$ and the undeformed $T_c(1,1)$ as a function of the deformation parameters $q_1$ and $q_2$ for the case $(q_1, q_2) \leq 1$.

**Figure 6.** The ratio $T_c(q_1, q_2)/T_c(1,1)$ of the $(q_1, q_2)$-deformed critical temperature $T_c(q_1, q_2)$ and the undeformed $T_c(1,1)$ as a function of the deformation parameters $q_1$ and $q_2$ for the case $(q_1, q_2) \geq 1$.

**Figure 7.** The specific heat $C_V/Nk$ as a function of $T/T_c(q_1, q_2)$ for various values of the deformation parameters $q_2$ and $q_1 \leq 1$.

**Figure 8.** The specific heat $C_V/Nk$ as a function of $T/T_c(q_1,q_2)$ for various values of the deformation parameters $q_2$ and $q_1 \geq 1$.

**Figure 9.** The gap in the specific heat $\Delta C_V/Nk$ at the critical temperature $T_c(q_1,q_2)$ as a function of the deformation parameters $q_1$ and $q_2$ for the case $(q_1,q_2) \leq 1$.

**Figure 10.** The gap in the specific heat $\Delta C_V/Nk$ at the critical temperature $T_c(q_1,q_2)$ as a function of the deformation parameters $q_1$ and $q_2$ for the case $(q_1,q_2) \geq 1$.

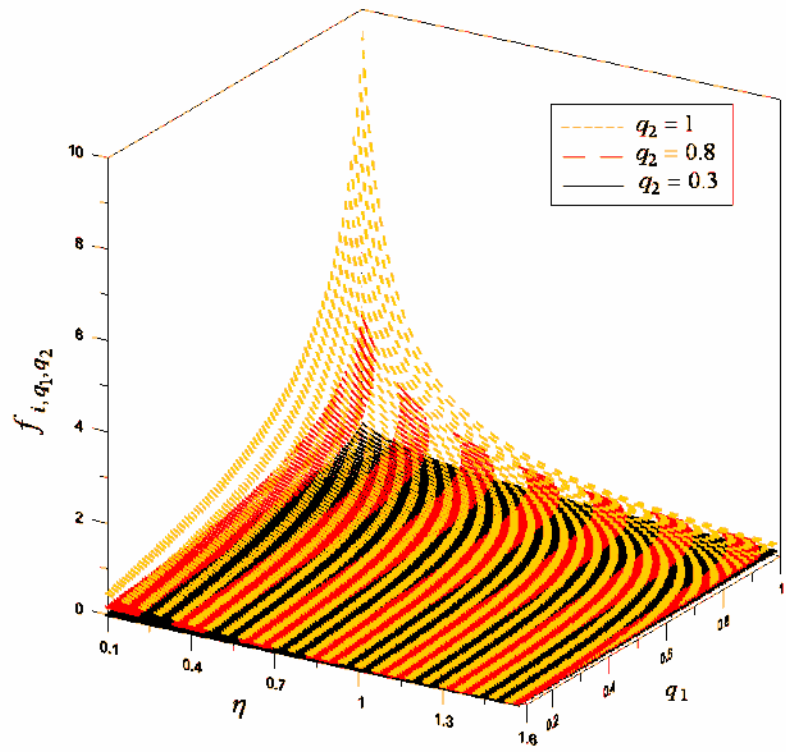

**Figure 1.**

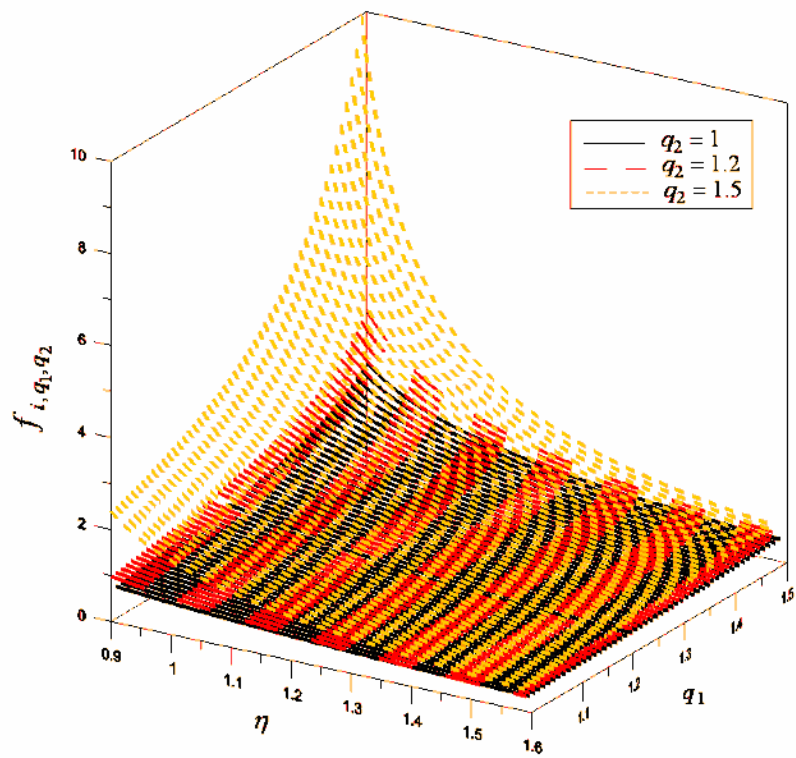

**Figure 2.**

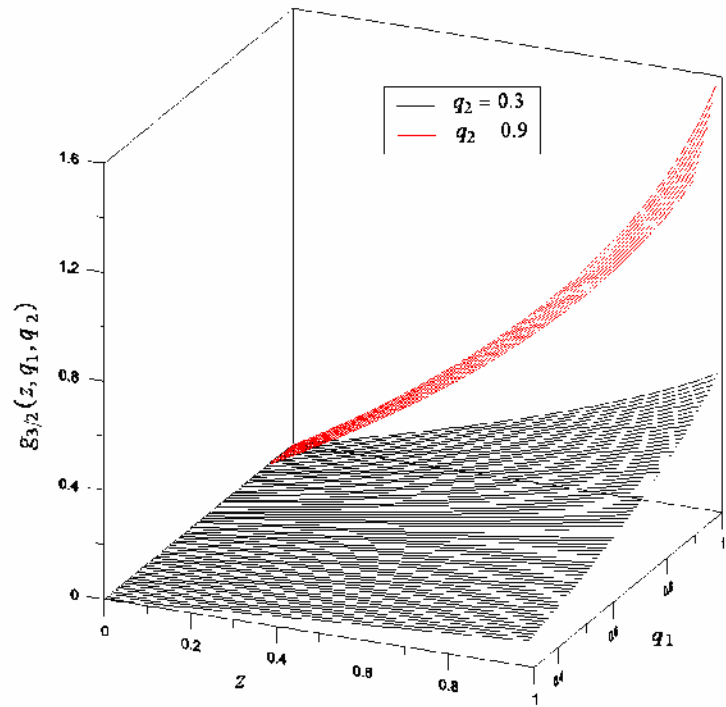

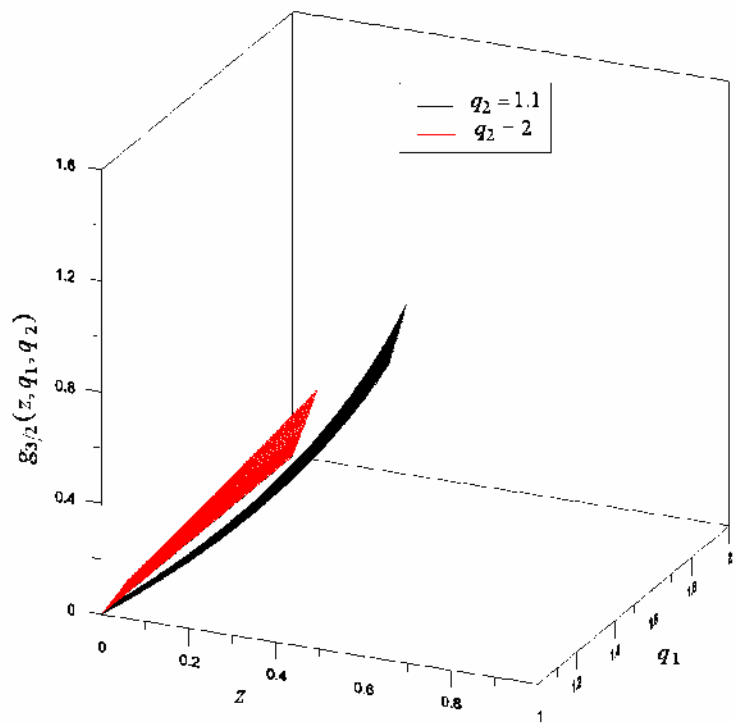

**Figure 3.**

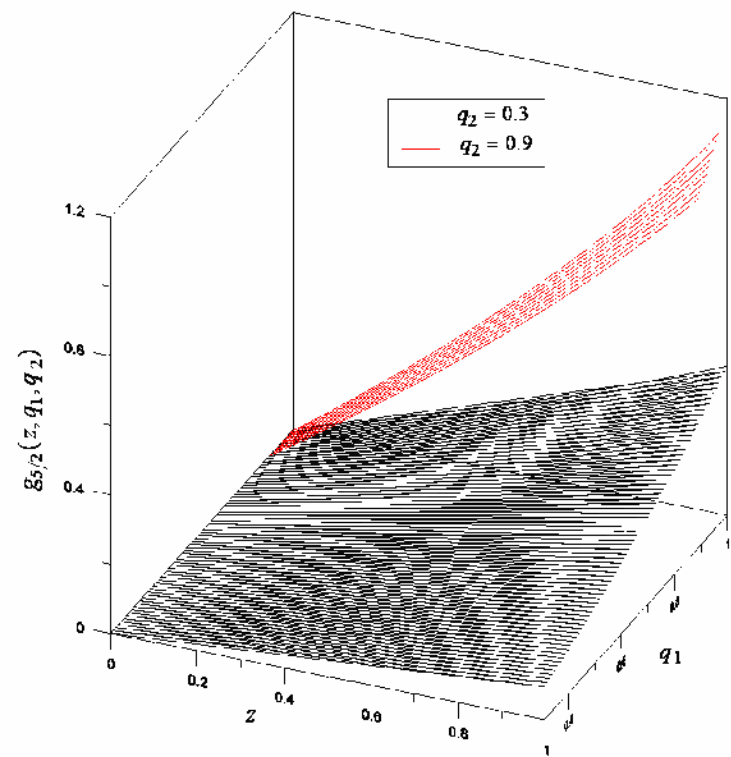

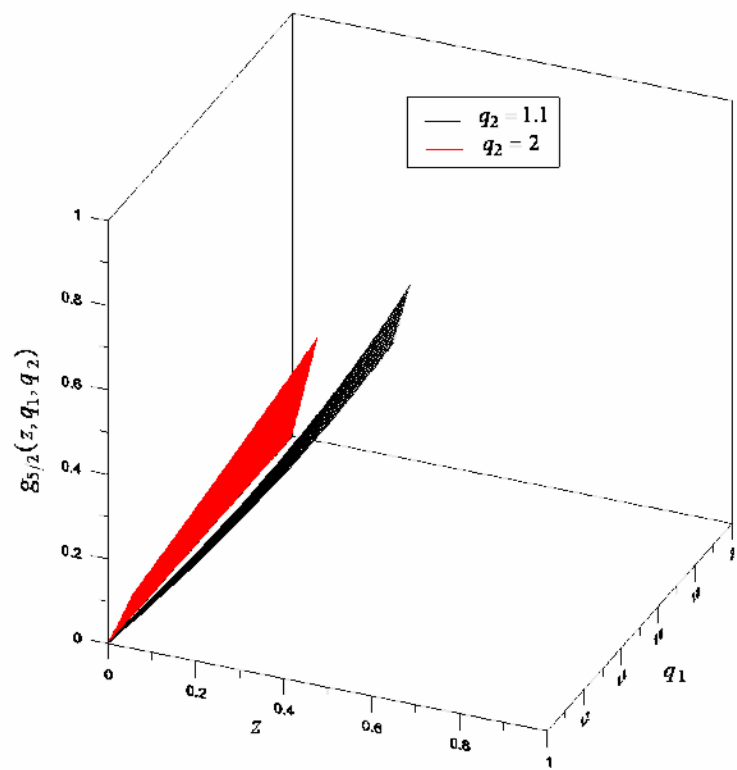

**Figure 4.**

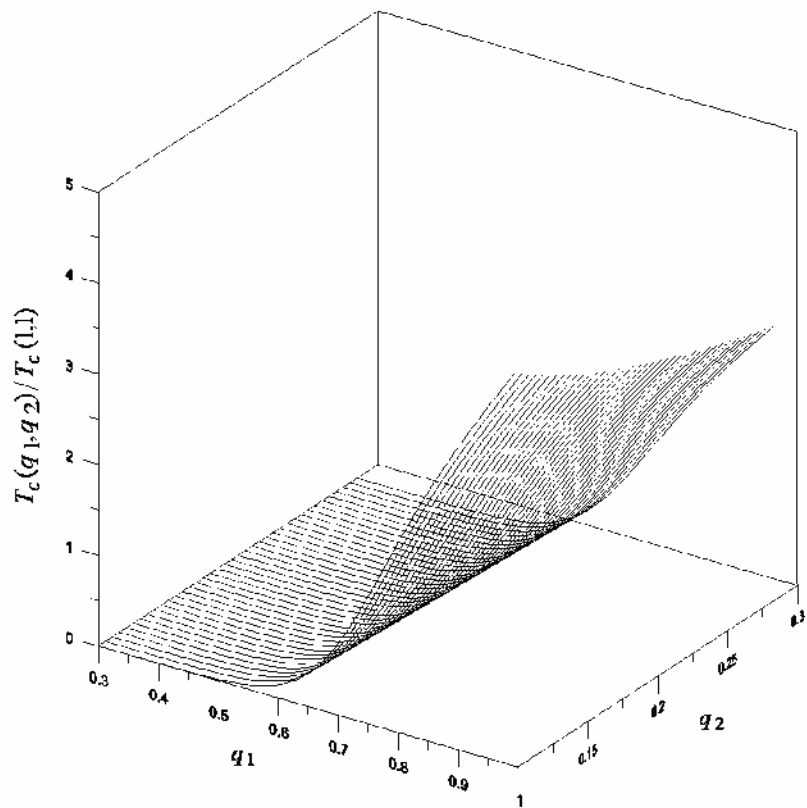

**Figure 5.**

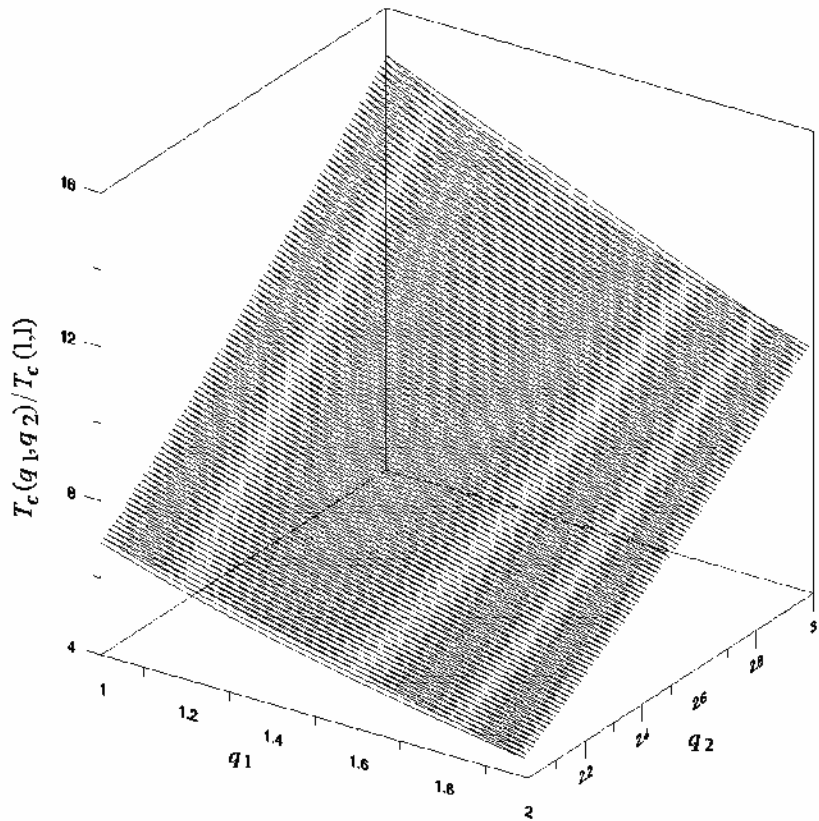

**Figure 6.**

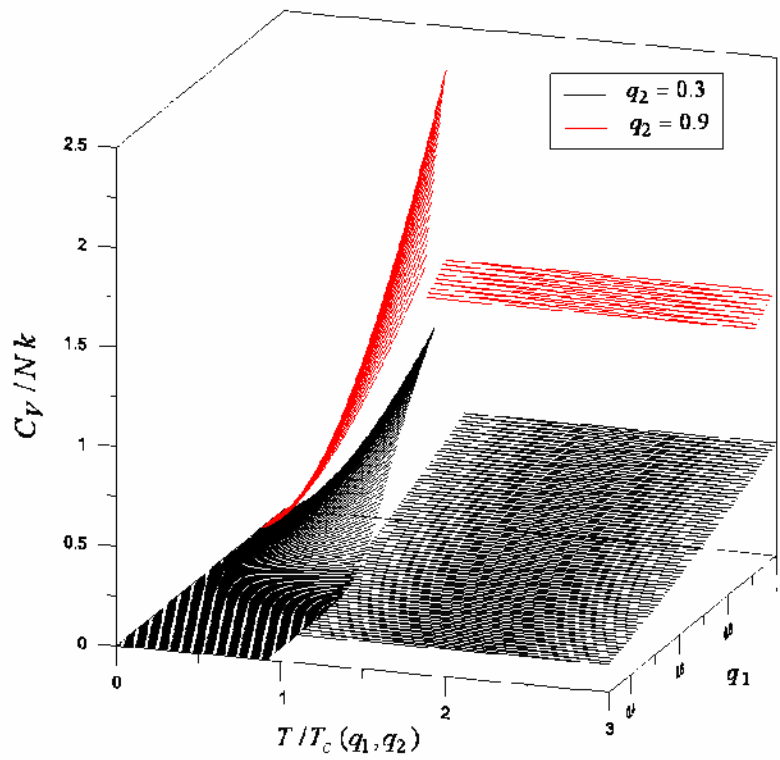

**Figure 7.**

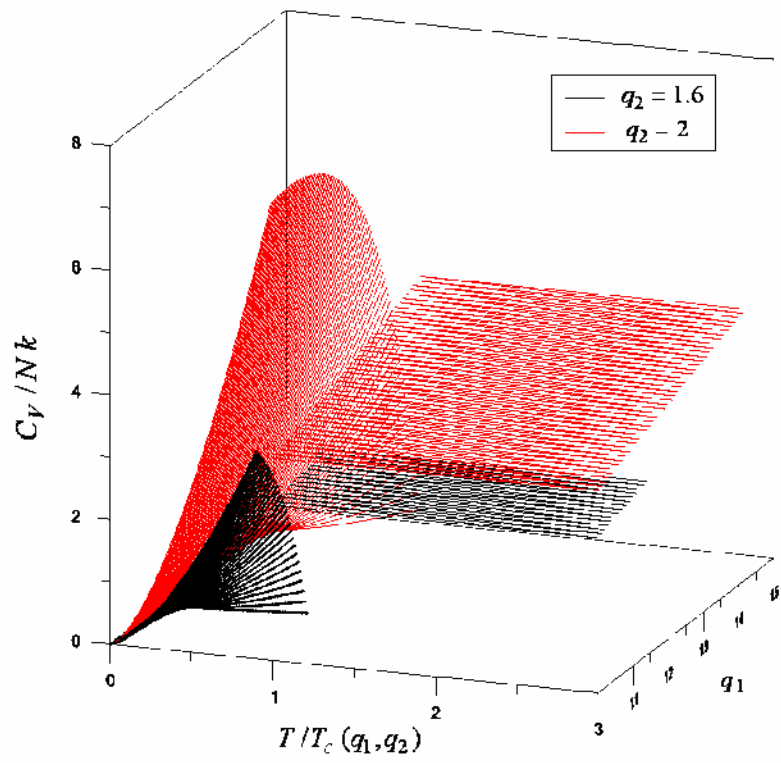

**Figure 8.**

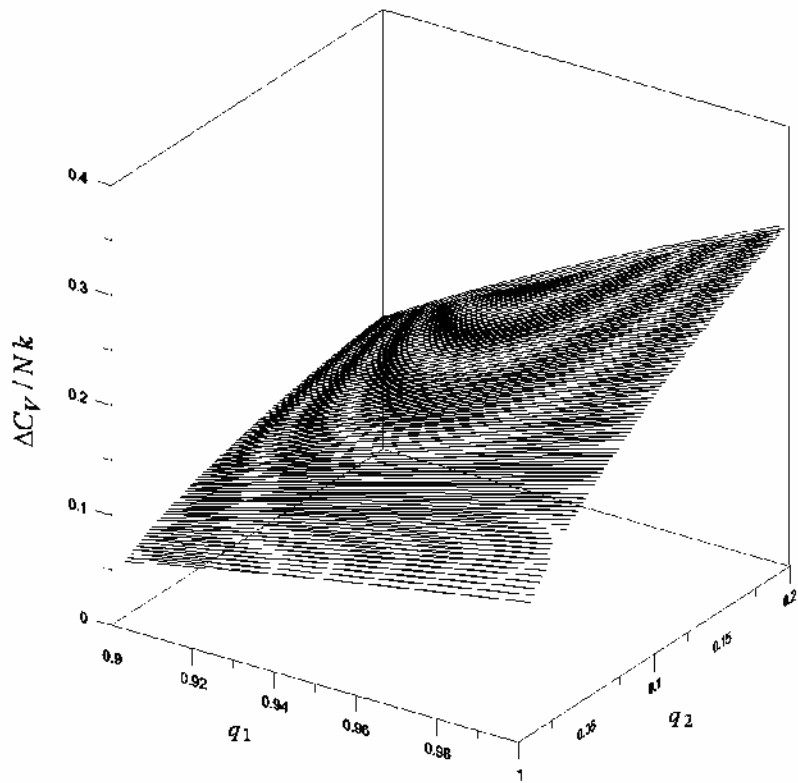

**Figure 9.**

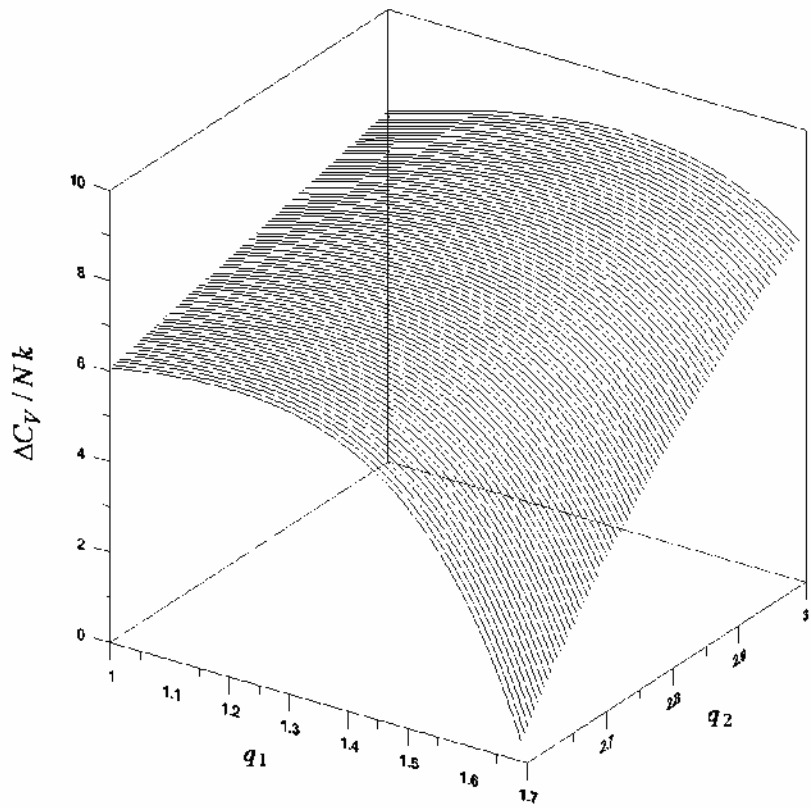

**Figure 10.**